\documentclass[final]{siamltex}

\def\H{{\mathscr{H}}}
\def\F{{\mathscr{F}}}

\def\Heaviside{{\textrm{H}}}

\usepackage{amsfonts, mathrsfs, amsmath}
\usepackage{graphicx}

\title{Stieltjes integral theorem \& the Hilbert Transform}

\author{Luisiana Xavier Cundin\thanks{Contractor: Conceptual Mindworks, Inc., 9830 Colonnade Blvd \#377, San Antonio, TX 78230 ({\tt luisiana.cundin.ctr@amedd.army.mil}).} \and Norman Barsalou\thanks{Naval Medical Research Unit - San Antonio (NAMRU-SA), Fort Sam Houston, TX 78234 ({\tt norman.barsalou@amedd.army.mil}).}}

\begin{document}

\maketitle

\begin{abstract}
Stieltjes integral theorem is more commonly known by the phrase 'integration by parts' and enables rearrangement of an otherwise intractable integral to a more amenable form; often permitting completion of an integral in closed form. Applying Stieltjes integral theorem to the Hilbert transform yields an alternate integral definition, which is homeomorphic and exhibits accelerated computation. By virtue of the convolution theorem, the integral is mapped to Fourier image space and delineates requirements for the inverse Fourier transform, also, these requirements reveal the underlying reason for increased computational speed. Lastly, an alternative to Cauchy's integral formula is deduced by extending the line integral with logarithmic kernel into the complex domain. 
\end{abstract}



\pagestyle{myheadings}
\thispagestyle{plain}
\markboth{L. X. Cundin \& N. Barsalou}
{Stieltjes Integral Theorem \& The Hilbert Transform}

\section{Prop{\ae}deutics}
The Hilbert transform is not uniquely defined in the literature and differs by a minus sign; to avoid confusion, the Hilbert transform is defined in Definition (\ref{hilbertdefinition}) and was adapted from R. N. Bracewell, also, the descriptor 'of the first form' will become evident later in this essay \cite{Bracewell}.
\begin{definition}[Hilbert transform of the first form]
The classical definition for the Hilbert transform ($\H$) of a function f(x) on domain $\mathcal{D}$ is defined to be the following definite integral: 
\begin{equation}\label{hilbertdefinition}
\H\Big\{f(x)\Big\}=-\frac{\mathcal{P}}{\pi}\int_{-\infty}^{\infty}{\frac{f(x^\prime)}{x-x^\prime}dx^\prime}
\end{equation}
where, the Cauchy principal value $\mathcal{P}$ is taken at all discontinuities.
\end{definition}

The Hilbert kernel, $(x-x^\prime)^{-1}$, experiences a discontinuity for every case where $x^\prime=x$; this property classifies the Hilbert transform as an \emph{improper} integral \cite{Widder}. The function $f(x)$ may possess additional undefined points within the domain; regardless, being an improper integral bars treating the Hilbert transform as a Riemann integral. For an arbitrary function $f(x)$, integration often necessitates expanding into the complex domain; whereupon, Cauchy's integral theorem proves indispensable, thus, $\mathcal{P}$ represents the \emph{Cauchy principal value}. 

For an $n$-dimensional Euclidean space, the Hilbert transform can be applied along each dimension, $\mathcal{D}\subseteq\mathbb{R}^n$, provided the integrals exist. The Hilbert transform can also be applied to complex domains, $\mathcal{D}\subseteq\mathbb{C}^n$, once again, assuming integral convergence. If a finite number of poles exist with no accumulation points in the complex domain for complex function $f(z)\in\mathbb{C}^n$, then the function is said to be meromorphic \cite{Davies}. A closed Jordan curve removing all poles from the domain forms a piecewise holomorphic function that represents the original meromorphic function \cite{Davies}. Holomorphic functions are continuous, analytic functions. Cauchy's integral theorem may be applied to holomorphic functions to evaluate the Hilbert transform of the function $f(z)$ in the upper half of the complex plane. 

Consider a real-valued constant function $f(x)$ over the interval $a\leq x\leq b$, the resulting integral is null, see equation (\ref{nullresult}). If interval $(0,\delta)$ were deducted from the interval of integration for the Hilbert kernel, the value of the integral is $-\infty$, no matter how small $\delta$ is made \cite{Widder}. Direct numerical integration of an arbitrary function $f(x)$ must judiciously handle the symmetric requirements in the neighborhood of any singularity within the domain; because of this behavior, the Hilbert transform is unstable and difficult to integrate numerically \cite{SmithLyness}. 
\begin{equation}\label{nullresult}
 \mathcal{P}=\lim_{\epsilon\rightarrow 0}\left[\int_a^{-\epsilon}{\frac{dx}{x}}+\int_{\epsilon}^b{\frac{dx}{x}}\right]=0
\end{equation}

The Hilbert transform as a convolution integral is shown in equation (\ref{hilconv}), where an asterisk ($\ast$) represents convolution. The Fourier transform ($\F$) of the Hilbert kernel is the sign function, written as $\hbox{sgn}(s)$; in addition, the function is multiplied by the imaginary number, $i=\sqrt{-1}$. Fourier theory states the convolution of two functions is equivalent to multiplying the Fourier image of each function in Fourier image space. Taking advantage of the convolution theorem, the Hilbert transform is mapped to Fourier image space as the multiplication of the sign function times the Fourier image of the function $f(x)$, finally, the inverse Fourier transform ($\F^{-1}$) is applied to return to the original domain \cite{Bracewell}. 
\begin{equation}\label{hilconv}
\H\Big\{f(x)\Big\}=-\frac{1}{\pi x}\ast f(x)=\F^{-1}\Big\{i\ \hbox{sgn}(s)\F\Big\{f(x)\Big\}\Big\}
\end{equation}

The Fourier image of the Hilbert transform replaces the numerical instability of the Hilbert kernel by the unbounded sign function. Depending upon the nature of the function $f(x)$, \emph{Shannon's sampling theorem} may or may not be easily satisfied \cite{Shannon}. Because the sign function is unbounded, all discrete transforms risk numeric error associated with \emph{under-sampling}.  

The function $f(x)$ and the corresponding Hilbert transform of $f(x)$ comprise a pair of functions referred to in unison as a Hilbert pair. In combination, the Hilbert pair form what is generally known as an 'analytic signal' or 'strong analytic signal'; the pair of functions are heavily relied upon in signal processing and \emph{causal} physical theory, e.g. electromagnetic theory supports the well-known Kramers-Kr\"{o}nig relations \cite{Landau}. Each function of the Hilbert pair represents an orthogonal geometric trajectory and in tandem form a set of functions completely covering a Riemann manifold \cite{Lee}.  

\section{Assembly}
It is common practice to apply, then immediately reverse an operation to find a new mathematical identity; such is Stieltjes integral theorem. Stieltjes integral theorem is more commonly known as 'integration by parts' and is stated in Theorem (\ref{inbyparts4}), which has been adapted from D. V. Widder's codification \cite{Widder}. Integration by parts is frequently employed to transform an otherwise intractable integral to a form that is more amenable to known techniques of integration. For the purposes of this essay, let the first and second instances of integration in Stieltjes integral theorem be referred to as ''the first form'' and ''the second form'', respectively.   
\begin{theorem}[Stieltjes integral theorem]\label{inbyparts4} If integrand function $f(x)$ and integrator function $\alpha(x)$ are continuous on the interval (a, b), then the sum of integration for the first and second form equals the limit sum of the product $f(x)\alpha(x)$.
\begin{equation}\label{stieltjes}
 \int_a^b{f(x)d\alpha(x)}+\int_a^b{\alpha(x)df(x)}=f(b)\alpha(b)-f(a)\alpha(a)
\end{equation}
\end{theorem}

\emph{Proof}: Take the derivative of the product $f(x)\alpha(x)$ immediately followed by integration, i.e. $\int{d(f(x)\alpha(x))}$. Applying the product rule for derivatives yields two terms, whereupon integration distributes and yields the sum of two integrals, finally, the limit sum is a consequence of the \emph{Fundamental Theorem of Calculus} \cite{Widder}. 

The Hilbert transform, Definition (\ref{hilbertdefinition}), can be identified as the first integral form in Stieltjes integral formula. To find the Hilbert transform of the second form, we first identify the integrator function $\alpha(x)$ to be the indefinite integral of the Hilbert kernel, which is the natural logarithmic function and leads to the following intermediary form:
\begin{equation*}
\H\Big\{f(x)\Big\}=\frac{\mathcal{P}}{\pi}\int_{-\infty}^{\infty}{f(x^\prime)d\ln(x-x^\prime)}
\end{equation*}

Reversing the role the integrand and integrator functions play yields the Hilbert transform of the second form, see Definition (\ref{hilbertdefinition2}), where the convolution form of the integral is also shown. The definition consist of the derivative of the function $f(x)$ multiplied by a shifted logarithmic kernel, then integrated over the entire real line. 
\begin{definition}[Hilbert transform of the second form] The Hilbert transform of the second form ($\H_{(2)}$) is defined by the following definite integral over the entire real line:
\begin{equation}\label{hilbertdefinition2}
\H_{(2)}\Big\{f(x)\Big\}=\frac{\mathcal{P}}{\pi}\int_{-\infty}^{\infty}{\ln(x-x^\prime)\frac{df(x^\prime)}{dx^\prime}dx^\prime}=\frac{1}{\pi}\ln(x)\ast df(x)
\end{equation}
where, the Cauchy principal value $\mathcal{P}$ is taken at all discontinuities.
\end{definition}

An interesting property of the convolution theorem arises when conjoined with the derivative theorem, namely, if one attempts to take the derivative of the entire convolution integral, one is free to operate the derivative on one function singly and then convolve with the other function, which has been untouched by any previous operation; the choice of which function to operate the derivative is completely arbitrary. 
\begin{theorem}[Derivative of a Convolution Integral]\label{derconv} The derivative of a convolution is equal to the convolution of either function with the derivative of the other \cite{Bracewell}.
\begin{equation}
 (f\ast g)^\prime=f^\prime\ast g=f\ast g^\prime
\end{equation}
where prime signifies a first order derivative.
\end{theorem}

Theorem (\ref{derconv}) provides an alternate proof for the Hilbert transform of the second form. Beginning with the convolution form of the Hilbert transform of the first form, the derivative is extracted, then applied to the function $f(x)$, also, the derivative is signified by prime rather than the italicized letter \textit{d}. The omission of a minus sign is intentional; retention of the minus sign leads to the negative of the result shown below. The absence of the minus sign indicates the Hilbert transform of the second form is, in fact, equal to the negative of the Hilbert transform of the first form.
\begin{equation}\label{convproof}
 \frac{1}{\pi x}\ast f(x)=\frac{1}{\pi}\Big(\ln(x)\ast f(x)\Big)^\prime=\frac{1}{\pi}\ln(x)\ast f^\prime(x)
\end{equation}

The Hilbert transform of the second form can be mapped to the product of two Fourier image functions in Fourier space by invoking the convolution theorem. It is required to know the Fourier transform of the natural logarithmic function to complete this transformation and this affords another opportunity to implement Stieltjes integral theorem. The symmetric Fourier transform of the natural logarithmic function is stated on the lefthand side of equation (\ref{fstiel}). Employing Theorem (\ref{inbyparts4}), the integral involving the logarithm and Fourier kernel can be rearranged and equated to the righthand side of equation (\ref{fstiel}).
\begin{equation}\label{fstiel}
 \int_{-\infty}^{\infty}{\ln(x)e^{-2\pi ixs}dx}=\frac{1}{2\pi is}\int_{-\infty}^{\infty}{\frac{1}{x}e^{-2\pi ixs}dx}-\frac{1}{2\pi is}\ln(x)e^{-2\pi ixs}\Bigg|_{-\infty}^{\infty}
\end{equation}

The integral on the righthand side of equation (\ref{fstiel}) transforms $1/x$ to $-i\pi \textrm{sgn}(s)$. The limit sum of the product between the natural logarithmic function and the Fourier kernel would appear, at first glance, ambiguous; because, the Fourier kernel oscillates and does not converge at infinity. The ambiguity can be managed by considering either the limit inferior or limit superior of the functions involved.

Evaluating the limit sum gives rise to two terms: the first term is evaluated approaching the infinite point along the positive direction of the real line; the second approaches minus infinity from the negative direction. In the case of evaluating the logarithmic function at minus infinity, the \emph{principal value} of the natural logarithm is introduced, yielding the absolute and argument ($\arg$) of the independent variable; finally, substitution leads to the second and third terms found on the righthand side of equation (\ref{limitsum}). The limit superior ($\varlimsup$) of both the first and second terms cancel. The limit superior of the last term, the argument (arg) times the Fourier kernel, is equal to $i\pi$. 
\begin{align}\label{limitsum}
 \ln(x)e^{-2\pi ixs}\Bigg|_{-\infty}^{\infty}=&\varlimsup_{x\rightarrow\infty}\Big\{\ln(x)e^{-2\pi ixs}\Big\}-\varlimsup_{x\rightarrow -\infty}\Big\{\ln|x|e^{-2\pi ixs}\Big\}-\\\nonumber 
&\varlimsup_{x\rightarrow -\infty}\Big\{i\arg(x)e^{-2\pi ixs}\Big\} \Rightarrow -i\pi
\end{align}

The sign of the limit involving the argument multiplied by the Fourier kernel is ambiguous, for the limit inferior would return $-i\pi$; hence, the solution to the limit sum is, in general, $\pm i\pi$. Collecting results, the Fourier transform of the natural logarithmic function can be displayed succinctly in equation (\ref{flog}), where the double arrow signifies a simplification of the algebraic formula after realizing the following equivalent relationship: $\hbox{sgn}(s)/s\equiv 1/|s|$.  
\begin{equation}\label{flog}
 \F\Big\{\ln(x)\Big\}=-\frac{i \pi\textrm{sgn}(s)}{2\pi is}\mp\frac{i\pi}{2\pi is}\Rightarrow-\frac{1}{2}\left(\frac{1}{|s|}\pm\frac{1}{s}\right)
\end{equation}

Using equation (\ref{flog}), the Fourier image of the Hilbert transform of the second form involves the Fourier transform of the natural logarithmic function and the Fourier image of the first derivative of the function $f(x)$, stated in equation (\ref{four1}). Since the Fourier image of the derivative of a function is equal to $2\pi i$ times the Fourier image of that function $f(x)$, we can swiftly dispense with the troubling prospect of taking the derivative of what could be a discrete set meant to represent the continuous function $f(x)$. Multiplication by $2\pi is$ and subsequent algebraic simplification leads to the last equality, specifically, right of the double arrow in the following:
\begin{equation}\label{four1}
\F\Big\{\H_{(2)}\Big\{f(x)\Big\}\Big\}=\frac{-1}{2\pi}\left(\frac{1}{|s|}\pm\frac{1}{s}\right)
\F\Big\{df(x)\Big\}\Rightarrow-i\left(\frac{s}{|s|}\pm 1\right)
\F\Big\{f(x)\Big\}
\end{equation}

As the transform variable $s$ varies over all real numbers, $\{s|s\in\mathbb{R}\}$, the algebraic term $s/|s|+1$ takes on the value of 2 for all positive real numbers or zero for all negative real numbers; this can be compactly represented by twice Heaviside's step function, written as $2\Heaviside(s)$. Heaviside's step function evaluates to 1/2 at the origin and the algebraic term evaluates to twice unity at the origin, hence, we must add a suitable term to account for the additional value missing, namely, the product between the imaginary number, Dirac Delta function $\delta(s)$ and the Fourier transform of the function $f(x)$. This allows further simplification of equation (\ref{four1}) and leads to equation (\ref{casessoln}). 
\begin{equation}\label{casessoln}
 \F\Big\{\H_{(2)}\Big\{f(x)\Big\}\Big\}=
\begin{cases}
 \Big(-2i \Heaviside(s)-i\delta(s)\Big)\F\Big\{f(x)\Big\}, \text{for }s/|s|+1\\
\\
 \Big(2i \Heaviside(-s)+i\delta(s)\Big)\F\Big\{f(x)\Big\}, \text{for }s/|s|-1
\end{cases}
\end{equation}

The Hilbert transform of the second form replaces the sign function with, essentially, twice Heaviside's step function; effectively nullifying either the entire negative or positive real line, depending upon which form is employed. As a consequence, an alternate perception of the Hilbert transform is that the transform suppresses either all negative or positive frequencies of a signal; thus, either advancing or retarding the phase of a signal.  

The inverse Fourier transform operated on both cases found in equation (\ref{casessoln}) would retrieve the Hilbert transform of the second form in the original domain. To that end, consider that the inverse Fourier transform of Heaviside's step function equals the sum of the Hilbert kernel and the Dirac Delta function $\delta(x)$:
$$\F^{-1}\Big\{\Heaviside(s)\Big\}=\frac{i}{2\pi x}+\frac{\delta(x)}{2}$$

Applying the above identity for Heaviside's step function to both cases found in equation (\ref{casessoln}) produces the results found in equation (\ref{invsec}). The initial result states that the function $f(x)$ is to be convolved with the Hilbert kernel with either the addition or subtraction of that function convolved with Dirac's Delta function, finally, either addition or subtraction of the imaginary number times the function $f(x)$ itself. 
\begin{equation}\label{invsec}
 \H_{(2)}\Big\{f(x)\Big\}=
 \frac{1}{\pi x}\ast f(x)\pm i\delta(x)\ast f(x)\pm if(x)
\end{equation}

Further simplification is attained by realizing the convolution of any function with the Delta function equals that function alone, namely, $\delta(x)\ast f(x)=f(x)$. Therefore, the inverse Fourier transform of equation (\ref{casessoln}) can be reduced to equation (\ref{reduced}). The earlier indication that the Hilbert transform of the second form was somehow related to the Hilbert transform of the first form can now be stated explicitly as a Corollary.
\begin{corollary}[Equivalence of the first and second form]\label{equiv} The Hilbert transform of the second form is equal to the negative of the classical Hilbert transform definition with either the addition or subtraction of function $f(x)$ multiplied by $2i$.
\begin{equation}\label{reduced}
 \H_{(2)}\Big\{f(x)\Big\}\equiv -\H\Big\{f(x)\Big\}\pm 2if(x)
\end{equation}
where the argument of the logarithmic function is responsible for the ambiguity in sign.
\end{corollary}

Substituting Corollary (\ref{equiv}) for the definition of the Hilbert transform of the second form in Steiljtes integral formula, Theorem (\ref{inbyparts4}), reduces the relation to a statement concerning the infinite point. Evidently, the residue at the infinite point is equal to $2\pi if(x)$; because of the ambiguity of the logarithmic function, the sign of the product is either positive or negative.
\begin{corollary}[Principal value]\label{hilbertdefinition3} Substitution of Corollary (\ref{equiv}) reduces Stieltjes integral formula for the Hilbert transform to the following statement:
\begin{equation}
\begin{split}
\H\Big\{f(x)\Big\}+\H_{(2)}\Big\{f(x)\Big\}&=\\ \H\Big\{f(x)\Big\}-&\H\Big\{f(x)\Big\}\pm 2if(x)=\frac{1}{\pi}\ln(x-x^\prime)f(x^\prime)\Bigg|^\infty_{-\infty}\Rightarrow\\
&\ln(x-x^\prime)f(x^\prime)\Bigg|^\infty_{-\infty}=\pm 2\pi if(x)
\end{split}
\end{equation}
\end{corollary}

Finally, using Corollary (\ref{equiv}) and Definition (\ref{hilbertdefinition2}), it is possible to deduce an alternate form for \emph{Cauchy's integral formula}, which states the line integral around a shifted logarithmic kernel and the first derivative of a function $f(z)$ is equal to the function $f(z)$ without differentiation. 
\begin{lemma}[Line integral with logarithmic kernel] Let $\Gamma$ be a closed Jordan curve in domain $\mathcal{D}$, then the line integral of the derivative of function $f(z)$ with a shifted logarithmic kernel is equal to the function $f(z)$:
\begin{equation}\label{cauchy}
f(z)=-\frac{1}{2\pi i}\oint_{\Gamma}{\ln(z-z^\prime)\frac{df(z^\prime)}{dz^\prime}dz^\prime}
\end{equation}
\end{lemma}


An $n$-th differential form of equation (\ref{cauchy}) is possible, much like \emph{Cauchy's integral formula}; albeit, this is less interesting, for it only states the integral of $f^{(n+1)}(z)$ is equal to $f^{(n)}(z)$, thus, erasing only one degree of differentiation. In contrast, the derivative formula for \emph{Cauchy's integral formula} finds far more utility. It is further possible to apply successive derivatives to the logarithm, but this only returns a formula similar to \emph{Cauchy's integral formula}; thus, providing little difference overall. 

\section{Exhibition}
Real-world signals are inherently real-valued and more than likely represented by discrete sets. Because discrete sets are usually encountered, it is most natural to implement Fourier transforms to emulate the Hilbert transform. Accelerated computation of the inverse discrete Fourier transform is immediately obvious for a list of elements one half the length required by any other means or schema. Since the Hilbert transform of the second form nullifies one half of the real line, then the prospect of accelerated computation of the Hilbert transform exists.
\begin{table}[ht!]
\caption{Statistics for computational runtimes involving both forms of the Discrete Hilbert Transform (DHT), where the first form is the classical Hilbert transform and the second form is Definition (\ref{hilbertdefinition2}). The comparative percent increase for the 2nd-DHT algorithm is relative to that of the 1st-DHT algorithm. For an indicated trial number, the average runtime to complete the inverse discrete Fourier transform (DFT) is shown with accompanying standard deviation. The power $N$ determines the number of elements operated on by the forward DFT; the length of the list is equal to $N-1$ for the inverse discrete Fourier transformation in the case of the 2nd-DHT algorithm.} 

\begin{center} \footnotesize
\begin{tabular}{|c|l|l|r|r|r|} \hline  
Integral & Power & Trial & Percent & Mean runtime & Standard Deviation\\ 
form & $2^N$ & number & increase & (milliseconds) & (milliseconds) \\ \hline

 & \lower.3ex\hbox{$10$} & \lower.3ex\hbox{3000} &  
\lower.3ex\hbox{---} & \lower.3ex\hbox{2.367} & \lower.3ex\hbox{5.600} \\ 
  
\lower.3ex\hbox{\textit{1st-DHT}} & \lower.3ex\hbox{$12$} & \lower.3ex\hbox{3000} &  
\lower.3ex\hbox{---} & \lower.3ex\hbox{9.204} & \lower.3ex\hbox{104.800} \\ 

 & \lower.3ex\hbox{$12.5$} & \lower.3ex\hbox{300} &  
\lower.3ex\hbox{---}  & \lower.3ex\hbox{3.269} & \lower.3ex\hbox{0.770} \\

 & \lower.3ex\hbox{$18.5$} & \lower.3ex\hbox{300} &  
\lower.3ex\hbox{---}  & \lower.3ex\hbox{569.509} & \lower.3ex\hbox{42.930} \\ 

 & \lower.3ex\hbox{$20$} & \lower.3ex\hbox{300} &  
\lower.3ex\hbox{---}  & \lower.3ex\hbox{3283.840} & \lower.3ex\hbox{185.500} \\
\hline
   
\lower.3ex\hbox{} & \lower.3ex\hbox{$10$} & \lower.3ex\hbox{3000} &  
\lower.3ex\hbox{515} & \lower.3ex\hbox{0.385} & \lower.3ex\hbox{2.561} \\ 
  
\lower.3ex\hbox{\textit{2nd-DHT}} & \lower.3ex\hbox{$11$} & \lower.3ex\hbox{3000} &  
\lower.3ex\hbox{582} & \lower.3ex\hbox{1.472} & \lower.3ex\hbox{4.562} \\ 

 & \lower.3ex\hbox{$12.5$} & \lower.3ex\hbox{300} &  
\lower.3ex\hbox{1364}  & \lower.3ex\hbox{0.223} & \lower.3ex\hbox{0.417} \\

& \lower.3ex\hbox{$18.5$} & \lower.3ex\hbox{300} &  
\lower.3ex\hbox{1616} & \lower.3ex\hbox{33.180} & \lower.3ex\hbox{6.893} \\

 & \lower.3ex\hbox{$20$} & \lower.3ex\hbox{300} &  
\lower.3ex\hbox{347} & \lower.3ex\hbox{740.877} & \lower.3ex\hbox{60.865} \\ \hline

\end{tabular}
\end{center} 
\label{diffstats} 
\end{table}  

Inspecting Table (\ref{diffstats}), there is an immediate computational savings from halving the length of any given list of data points with regard to Fourier inversion; but, any apparent benefit swiftly diminishes as the length of the list increases. This can be explained by the fact that regardless of the algorithm used the difference in the number of operations required for the inversion becomes negligible to the total required number of operations as elements are increased; nevertheless, measured statistics clearly show a marked increase in all cases tested. 

Lists possessing a length equal to an integer power of two, generally speaking, fall under the class of algorithms referred to as the \emph{Fast Fourier Transform}. As evidenced by the data in Table (\ref{diffstats}), it is apparent that lists of length not equal to an integer power of two took less time to compute; an explanation of this behavior is impossible without being privy to internal proprietary knowledge of \emph{Mathematica's} algorithms. 

Because the internal workings of \emph{Mathematica's} \textbf{InverseFourier} function is not privy, it was decided to use the average of several successive runs to remove proprietary optimization methods from biasing any of the tests. Be it as it may, the percent increase in computational speed is relative and not dependent upon the specific workings of the algorithm employed nor computer language, hardware architecture, \&c. Regardless of the specific runtime for each length of list entertained, the relative computational increase in runtime was faster for the second Discrete Hilbert Transform algorithm (2nd-DHT) over the classical Discrete Hilbert Transform algorithm (1st-DHT).

The infinity norm of the difference between numeric output from both algorithms were within machine error and determined both numerically identical, specifically, $||\epsilon||_{\infty}< 15.9546$-logarithm base ten scale; furthermore, this was true for any test function employed. Runtimes were calculated using \emph{Mathematica's} built-in \textbf{Timing} function, intended for this purpose. All numeric calculations were written and executed using Mathematica version 5.2 software, from Wolfram Research, installed on an AMD Phenom\texttrademark\ Triple-Core Processor 2.30 GHz machine, with 2.00 GB RAM memory. 

\section{Epilog}
The Hilbert transform enjoys prominence in Harmonic theory and sundry uses spanning many disparate disciplines in both theoretical and engineering applications. With the aid of Stieltjes integral theorem, a new novel integral form is found for the Hilbert transform, dubbed 'the Hilbert transform of the second form' for a lack of a better term. The second integral form involves an integral over the entire real line with a logarithmic kernel, of which, this form of integral often arises in many engineering and physical applications \cite{Yan,Estrada}. 

Applying the Hilbert transform generates the Harmonic conjugate of \emph{causal} real-valued signals. In unison, the Hilbert pair constitutes an analytic signal and embodies all potential information contained in the measurement, i.e. amplitude, phase and frequency. Many well-behaved continuous functions are susceptible to closed form transforms; albeit, such signals are ideal and serve mostly theory. In general, measured signals are represented by a discrete set of data points; furthermore, it is to this sort of data representation that the Hilbert transform is more likely brought to bear and the underlying continuous functions rarely gleamed. 

It happens that many numerical methods exist to evaluate the Hilbert transform; yet, given the discrete nature, it is natural and most efficient to implement Fourier transforms in exercise. Benchmark tests show computational savings for the Discrete Hilbert Transform of the second form; furthermore, savings are solely due to the length of the list being truncated before the inverse Fourier transform is applied. 

Finally, by extending the domain to include a closed Jordan curve within the complex domain, the formula for the Hilbert transform of the second form may be cast as a contour integral; thus, providing an equivalent formula for \emph{Cauchy's integral formula}, but in this case involving a logarithmic kernel. The utility of an alternate \emph{Cauchy integral formula} becomes evident for integrals with logarithmic kernels; either enabling direct integration or providing a transform to an integral explicitly not involving a logarithm \cite{Yan,Estrada}.

\section*{Acknowledgments}
The author, L. X. Cundin, would like to acknowledge Professors Ashok Puri, George Ioup, Juliette Ioup, Curtis L. Outlaw, Pratap Puri and all the faculty of the University of New Orleans; without which, impoverished youth throughout the City of New Orleans would have \emph{no hope} of matriculation. 

I am a military service member (or employee of the U.S. Government). This work was prepared as part of my official duties. Title 17 U.S.C. §105 provides that 'Copyright protection under this title is not available for any work of the United States Government.' Title 17 U.S.C. §101 defines a U.S. Government work as a work prepared by a military service member or employee of the U.S. Government as part of that person's official duties.

The views, opinions and/or finding contained in this report are those of the authors and should not be construed as an official Department of the Navy, Air Force and Defense position, policy or decision unless so designated by other documentation. Trade names of materials and/or products of commercial or nongovernmental organizations are cited as needed for precision. These citations do not constitute official endorsement or approval of the use of such commercial materials and/or products.

This work was funded by NAMRU-SA work unit number 1LA403.

\bibliography{references}   
\bibliographystyle{plain} 

\end{document}